\documentclass[prd,twocolumn]{revtex4}

\newcommand{\be}{\begin{equation}} 
\newcommand{\ee}{\end{equation}}
\newcommand{\scp}{{\cal I}^+}
\newcommand{\scm}{{\cal I}^-}
\newcommand{\calg}{{\cal G}}

\begin{document}

\title{Dwelling on de Sitter}
\thanks{EFI-02-82, hep-th/0208018}

\author{Bruno Carneiro da Cunha}
\thanks{bcunha@theory.uchicago.edu}

\affiliation{Enrico Fermi Institute and Department of Physics,
The University of Chicago, 5640 South Ellis Avenue, Chicago Illinois
60637}

\begin{abstract}

A careful reduction of the three-dimensional gravity to the Liouville
description is performed, where all gauge fixing and on-shell
conditions come from the definition of asymptotic de Sitter
spaces. The roles of both past and future infinities are discussed and
the conditions space-time evolution imposes on both Liouville fields
are explicited. Space-times which correspond to non-equivalent
profiles of the Liouville field at $\scm$ and $\scp$ are shown to
exist. The qualitative implications of this for any tentative dual
theory are presented.   

\end{abstract}

\maketitle

\section{Introduction}

The importance of understanding quantum gravity in de Sitter space can
hardly be overstressed. Touching physical problems like the dynamics
of inflation and the eventual fate of the universe, the question is in
itself a necessary step towards a background-free formulation of the
quantum theory of gravity and quantum cosmology.

Not having a precise string theory background in
which to embed the space, our understanding relies on extrapolations
of ideas gathered in the study of black holes and anti de Sitter
(AdS) spaces. A particularly useful idea is holography
\cite{holo}, which found a realization for AdS backgrounds \cite{sw}. 
With it in mind, Strominger gave a prescription \cite{strominger} to
recover data pertaining the space-time based on a hypothetical dual
(holographic) theory. The arguments in support of this prescription
are numerous \cite{bms, dscft}, although few seem to tackle properties of dS
which do not parallel those of AdS.

The properties of dS which make the formulation of holography
particularly difficult are the existence of a horizon and the
compactness of spatial slices. The former poses problems for unitary
evolutions of Hilbert spaces assigned to Cauchy surfaces and the
latter makes unclear the procedure to define conserved
charges. Horizons are in fact already known to cause problems in AdS,
where the formulation of (perturbative) string theory in, say, BTZ
backgrounds is an open problem. 

Also, if one chooses to think about holography in a covariant way
\cite{bousso}, the boundary theory seems to live in two disjoint
codimension one surfaces, the past and future infinities, $\scm$
and $\scp$ respectively. Strominger gave some arguments in favor
of a reduction which would map states from one to the other. Those
were based on the initial value formulation which would determine the
final state. Whether this can be brought forward to the full quantum
theory is unclear.

The aim of this article is to give an alternative way of regarding
this map, at least for three dimensional spacetimes. The arguments
presented here infer that the mapping is a truncation. One can argue
that the full theory does live in both infinities, but some states
(which we will call static) on it corresponding to singularity-free
solutions can be ``projected'' to either infinity by a gauge
choice. We will also consider instances where such projection fails.

The paper is organized as follows. In section II a reduction from the
Chern-Simons formulation to Liouville theory is carried out. We choose
a different ``gauge fixing'' from previous work \cite{ck} in order to
keep the spacetime picture explicit. In section III we will be able to give
a simple interpretation for the Liouville field, and then state the
generic conditions under which an asymptotically de Sitter spacetime
will be described by it. In section IV we discuss the Strominger
mapping and problems arising from cosmological singularities. We then
conclude with a prospectus of what else classical de Sitter gravity
could say about holography. 

\section{Chern-Simons and Wess-Zumino formulations revisited}

Three-dimensional gravity has been the subject of extensive research
over the past two decades \cite{3dgr}. The Chern-Simons formulation
\cite{w,at} makes explict the role of local isometry
invariance and also the fact that the theory is locally
trivial. Sticking to the positive cosmological constant case, we can
write the action in terms of an $SL(2,C)$ connection, defined as
$A^i_a=\omega^i_a+i\ell^{-1}e^i_a$, where $e^i_a$ is the
dreibein and $\omega^i_a$ is the Lorentz dual to the spin
connection. Then the properties above are readily seen by direct
inspection of the action,
\be
S=\frac{\ell}{4\pi G}{\Im}\int_M \mbox{Tr}(A\wedge
dA+\frac{2}{3} A\wedge A\wedge A)+ B(A,\bar{A}) \label{action}
\ee
where $A=A^i\gamma_i$, $\bar{A}$ is the complex conjugate to $A$ and
$\Im$ denotes the imaginary part. $\gamma_i$ can be though of as a
representation of the $SL(2,C)$ algebra in three dimensions, satisfying
$2\mbox{Tr}[\gamma_i\gamma_j]=\eta_{ij}$ and
$2\mbox{Tr}[\gamma_i\gamma_j\gamma_k]=\epsilon_{ijk}$. 
$B(A,\bar{A})$ is a boundary term
required to continue interior solutions to the causal boundary 
of spacetime. The properties of isometry invariance and local
triviality translate to trivial facts about 
Chern-Simons forms, respectively: i) they are gauge invariant and ii) their
variation with respect to $A$ is a closed form. The dynamics is then
determined by boundary data $B(A,\bar{A})$, namely, which
particular space the dynamic spacetime approaches asymptotically. In
the case at hand the sensible choice is to have the metric approach the de
Sitter metric at past and future infinity. 

In the spirit of classical general relativity, one should strive to give
a coordinate independent formulation of an asymptotic de Sitter
space. The main advantage of this approach is to make a clear
distinction between what is considered ``gauge choice'' in the
Chern-Simons language and what is considered ``on-shell''. Following
Shiromizu {\it et~al.} \cite{maeda} (after similar work done for AdS by
Hawking and by Ashtekhar and Magnon \cite{ham}), we will say that a
$n$-dimensional spacetime with metric $g_{ab}$ is {\em asymptotically
de Sitter} when 
\begin{enumerate}
\item {\em The causal boundary of space-time is
$\scp\cup\scm$ {\rm \cite{he}}.} Specifically, there are two functions
$\Omega^-$ and $\Omega^+$  which 
vanish at past and future infinities respectively, but not their
derivatives; also, the corresponding unphysical or fiducial metric
$\hat{g}^{\pm}_{ab}\equiv (\Omega^{\pm})^2 g_{ab}$ is regular there.
\item {\em The geometry is asymptotically trivial}: the space-time
$(M, g)$ is a solution of the equations $R_{ab}-\frac{1}{2}R 
g_{ab} + \Lambda g_{ab} = 8 \pi G T_{ab}$, where
$(\Omega^{\pm})^{-n+1}T^a_b$ has a smooth limit to the boundary.
\end{enumerate}

The first item above tell us that $\Omega$ is a non-singular
coordinate near either $\scm$ or $\scp$. Furthermore, its
gradient is normal to the boundary and hence timelike. So one is
naturally led to the {\it gauge choice} where the time function is
given by $t \sim \log \Omega^-$ near $\scm$. Whether or not one can
stick to this choice up to $\scp$ depends on the global issues like
the causal structure of spacetime and whether the function $t$ thus
defined can be extended globally. We will address these points in
section 3. Assuming that this is a good gauge choice
throughout the evolution of spacetime amounts to assuming that the
spatial slices will not undergo topology change. In fact, the
assumption already furnishes us with a Morse function, if we see
time-evolution as a cobordism. There is more to say about the
instances when such gauge fixing breaks down, but we will postpone it
for now.

The second item in the list relates to solving the equations
of motion near past and future infinity. Here we will be mostly
interested in the case of pure gravity, so $T_{ab}=0$ and the metric
is locally the de Sitter vacuum. It is interesting to note that if one
takes holography as a property of quantum gravity, the latter will
be represented in de Sitter backgrounds as a transfer matrix
interpolating between asymptotic de Sitter spaces at past and future
infinity. The requirement that these spaces satisfy item 2 above
echoes the LSZ reduction between general Green functions of any quantum
field theory and its $S$ matrix elements. Since the latter are defined
for on-shell elements, it is unclear at this point how the
quantum fluctuations of the bulk fields would be encoded in the dual
theory. As it stands, even in the AdS/CFT these are encoded in a
complicated way in the boundary theory. Only gauge-invariant
observables have a clear-cut correspondence in terms of bulk fields.

In the following we will apply the conditions above to the metric at
$\scm$ (and as such we will drop the superscript.) In terms of the
unphysical or fiducial variables, the Einstein equations in $n$
dimensions read:
\begin{eqnarray}
\hat{R}_{ac}\!&\!\! + \!&\!\! 
(n-2)\frac{1}{\Omega}\hat{\nabla}_a\hat{\nabla}_c\Omega +
\hat{g}_{ac}\hat{g}^{de}\frac{1}{\Omega}
\hat{\nabla}_d\hat{\nabla}_e\Omega 
\nonumber \\
& & \!\! - (n-1)\hat{g}_{ac}\hat{g}^{de}\frac{1}{\Omega^2}
\hat{\nabla}_d\Omega\hat{\nabla}_e\Omega  = 
\frac{2\Lambda}{n-2}\frac{1}{\Omega^2} \hat{g}_{ac} \\
\nonumber
\end{eqnarray}
Since the unphysical Ricci tensor is regular, the terms
diverging with some power of $\Omega$ must vanish at
$\partial\hat{M}$. The term proportional to $\Omega^{-2}$ yields
\be
\hat{g}^{ab}\hat{\nabla}_a\Omega
\hat{\nabla}_b\Omega =
-\frac{2}{(n-2)(n-1)} \Lambda \mbox{  at }\scm,
\ee
back to three dimensions, this means that one can define a time-like
vector field, $\xi^a \equiv \ell g^{ab}\nabla_b\log \Omega$, with
$\Lambda=\ell^{-2}$, which becomes
normalized in the physical space as one approaches $\scm$. The
integral curves of this vector are parametrized by a time
function, which we will call $t\equiv \ell \log\Omega$. One
can actually use part of the gauge symmetry of $\Omega$ and eliminate
the order $\Omega$ corrections to $\hat{\xi}^a$ \cite{wald2}. 
With this provision, the subleading term implies:
\be
\hat{\nabla}_a\hat{\nabla}_c\Omega + \hat{g}_{ac}\hat{g}^{de}
\hat{\nabla}_d\hat{\nabla}_e\Omega = 0 \mbox{  at }\scm\label{gauge2}
\ee
which in turn means that $\hat{\nabla}_a\hat{\nabla}_c\Omega =
0$. On top of those, there is a remnant gauge
condition, which allows it to be multiplied by a generic non-vanishing
function of the transverse coordinates. Since the spatial slices are
topologically spheres, one can use up the reparametrization invariance
and write the spatial metric as conformally flat. The conformal factor
can then be absorbed by a redefinition of $t$. The fiducial metric
can then be considered flat.

It is a straighforward exercise to compute the physical dreibein and
spin-connection given their fiducial counterparts and the function $t$:
\begin{eqnarray}
e^i_a & = & e^{-\ell^{-1}t}\hat{e}^i_a  \\
\omega^i_a & = &\hat{\omega}_a^k-\frac{1}{\ell}{\epsilon_{ij}}^k e_t^i e_a^j
\label{spin}  
\end{eqnarray}
where $e_t^i=\xi^a e_a^i$. Using local Lorentz invariance, one can
write $e_t^i=\delta^i_0$ at $\scm$. Also, given that the fiducial
metric is flat at $\scm$, $\hat{\omega}_a^k=0$ there and then the
following component of the gauge field satisfies
\be
\left.
\begin{array}{c}
A^+_a\equiv A^1_a+iA^2_a=0 \\
\\
A^0_z=A^0_{\bar{z}}=0 
\end{array} \right\} \mbox{ at }\scm
\label{onshell}
\ee
and that we can choose
\be
A^0_t=\frac{i}{\ell}
\label{gauge}
\ee
as a gauge fixing for (\ref{action}). As said before, we will see
(\ref{onshell}) as the on-shell condition, to be imposed at $\scm$,
and (\ref{gauge}) as a gauge choice.

The variation of the the action (\ref{action}) gives \cite{ms, emss}: 
\be
\delta S=\frac{k}{8\pi}\Im\left[\int_{\partial M}\!\!\!\! \mbox{Tr}(\delta
A\wedge A)+2\int_M\!\! \mbox{Tr}(\delta A \wedge F)\right]+\delta B
\label{variation} 
\ee
where $F=dA+A\wedge A$ and the ``level'' of the model is
$k=\frac{\ell}{2G}$. One sees that the stationary points of the
action correspond to flat connections. The allowed solutions are those
which can be continued to the boundary, {\it i.~e.}, whose boundary
variation vanishes. This of course depends on the exact form of
$B$. With our conditions (\ref{onshell}) and (\ref{gauge}) above, we
can see that the boundary term in (\ref{variation}) 
\be
\mbox{Tr}(\delta A\wedge A)=-\frac{1}{2}\delta A^0 \wedge A^0 +
\delta A^- \wedge A^+ + \delta A^+ \wedge A^- 
\ee
vanishes identically for connections giving rise to asymptotically de
Sitter spaces. At $\scm$ we can see it directly from the conditions
(\ref{onshell}). At $\scp$ one can parallel the argument given above
to find that the conditions (\ref{onshell}) hold with $A^-$ instead of
$A^+$. This comes about because we will need to define the time
function as $t=-\ell \log \Omega^+$ in order to have $\scp$ at
$t=\infty$. Then, by setting $B=0$, we make sure that only
connections which give rise to asymptotically de Sitter 
spaces are singled out. 

Considering the on-shell and gauge fixing conditions, one may ask what
is the resulting action after these are enforced in
(\ref{action}). The discussion above implies that 
$A^i_t$ is not a dynamical variable but rather is a Lagrange
multiplier. In fact, decomposing $d=dt 
\frac{\partial}{\partial t} + \tilde{d}$ and $A=A_{t}+\tilde{A}$, we
can write the action as
\begin{eqnarray}
S\!&\!\! = \!&\!\! \frac{k}{8\pi}\Im\left\{ 
\int_{M}\mbox{Tr}(\tilde{A}\wedge
\dot{\tilde{A}}\wedge dt)+
2\int_M\mbox{Tr}[A_{t}\wedge\tilde{F}] \right.
\nonumber \\
& & \;\;\;\;\;\;\;\;\;\;\;\;\;
\left. -\int_M \mbox{Tr}[\tilde{d}(\tilde{A}\wedge
A_{t})]\right\}
\label{gauge-fixed};
\end{eqnarray}
the last term vanishes due to the choice of coordinates. Now seeing
$A_{t}$ 
as a Lagrange multiplier, we can integrate over it and then enforce the
constraint that $\tilde{F}=\tilde{d}\tilde{A}+\tilde{A}\wedge
\tilde{A}=0$. For simply-connected manifolds, the solution of the
constraint is $\tilde{A}=-\tilde{d}\calg\calg^{-1}$ with $\calg$ an
$SL(2,C)$-valued function of the manifold. If the spatial slices have
punctures, we will have to deal with holonomies of $\calg$. We will
postpone this discussion to the next section. The on-shell condition
now translates into 
$A_t = -\dot{\calg}\calg^{-1}=i\ell^{-1}\gamma_0$, meaning that we can write
$\calg = e^{-i\ell^{-1}t\gamma_0}g(z,\bar{z})$, with $g$ independent of $t$.

Changing variables in (\ref{gauge-fixed}), and using the fact that the
resulting action depends only on the boundary values of $\calg$, one
arrives at: 
\hfuzz 5pt
\be
\!\!S[g]\!=\!-\frac{k}{4\pi}\Im\left\{\! \int\!\mbox{Tr}(g^{-1}\partial
gg^{-1}\bar{\partial} g)d^2x\!
+\!\frac{i}{6}\!\int_{M}\!\!\!\mbox{Tr}[(g^{-1}d g)^{\wedge
3}]\!\right\} \label{action2}
\ee
\hfuzz .1pt
where we used $d^2x=\frac{i}{2}dz\wedge d\bar{z}$ as the measure of the
boundary. 
One notes that $g$ actually gives information about the
spatial slices of the metric, or the fiducial metric since with it one can
recover the physical metric at $\scm$ by the conditions (\ref{onshell}) and
(\ref{gauge}). It is also clear by those conditions that not all $g$
will give rise to an acceptable metric. First and foremost, the metric
on the sphere has to be conformally flat, that is to say, we have already
made a choice for the K\"ahler structure on the spatial
slices. However, by inspecting the form of $A$ near $\scm$,
\be
A=\frac{i}{\ell}dt\gamma_0-e^{-i\ell^{-1}t\gamma_0} \tilde{d}gg^{-1}
e^{i{\ell}^{-1}t\gamma_0}
\ee
we see that $A^+$ vanishes at $t=-\infty$ independently of $g$ and
then it only makes sense to impose the flat condition through
$A^-$. By implementing the first-class constraint on (\ref{action2})
via the usual Lagrange multiplier, one arrives at the improved action:
\be
S_I[g,\Theta ]=S[g]+\frac{k}{2\pi}\Im \int d^2x\mbox{Tr}\left[\Theta
\left(\partial gg^{-1}-\frac{i}{\ell}\gamma_+\right)\right] 
\label{action3}
\ee
where $\Theta=\theta\gamma_-$ and $\theta$ is a scalar function. It is
well-known \cite{ds} that the action above has a gauge symmetry 
\be
g \rightarrow h(z,\bar{z})g,\;\;\;\;\;\;\;\;\Theta\rightarrow h\Theta
h^{-1}-dh h^{-1}\label{gauge3}
\ee
for $h$ in the lower triangular subgroup of $SL(2,C)$. The invariance
when $h$ is a function of $\bar{z}$ stems from the usual left-right
symmetry of (\ref{action2}). Parametrizing $g$ by the Gauss product:
\be
g=e^{\chi\gamma_-}e^{2i\rho\gamma_0}e^{\varphi\gamma_+}
\ee
we can write (\ref{action3}) as
\hfuzz 3pt
\be
\!\!S_I[g,\theta]\!=-\!\frac{k}{2\pi}\Im\int\!
d^2x\!\left[\partial\rho\bar{\partial}\rho+
e^{2\rho}\partial\varphi\bar{\partial}\chi+\theta\!\left(e^{2\rho}
\partial\varphi-\! \frac{i}{\ell}\right)\!\right]
\label{action4}
\ee
\hfuzz .1pt
The solution of the equation of motion for $\theta$ is then:
\be
e^{2\rho}=\frac{i}{\ell}(\partial\varphi)^{-1}\label{fixing2}
\ee
and one can use the gauge freedom
(\ref{gauge3}) to set
\be
\chi=-\frac{i\ell}{2}\frac{\partial^2\varphi}
{\partial\varphi}. \label{fixing3}
\ee
We will see later why this is a 
natural choice. The resulting action is then:
\be
S_I=\frac{k}{8\pi}\Im \int d^2x
\left[ \frac{\partial^2\varphi\partial\bar{\partial}\varphi}{(\partial
\varphi)^2}-2\frac{\partial^2\bar{\partial}\varphi}{\partial\varphi}
\right]
\label{action5}
\ee
which can be seen to be equivalent to the Liouville action if we write
its stress tensor, using the N\"other procedure:
\be
T=\frac{k}{4\pi}\left[\frac{\partial^3\varphi}{\partial\varphi}-
\frac{3}{2}\left(\frac{\partial^2\varphi}{\partial\varphi}\right)^2
\right] = \frac{k}{4\pi}\{\varphi ; z\}
\ee
or, in other words, the (unrestricted) stress tensor encodes the
Virasoro currents' Poisson brackets, which is a universal
characteristic of the Liouville field. In the equation above
$\{\varphi ; z\}$ denotes the Schwarzian derivative. This hardly comes
as a surprise 
since (\ref{action5}) is reminiscent of the natural geometrical action
for a particular coadjoint orbit of the Virasoro group (when the
symmetry group is real and compact and $b_0=0$ in the notation of 
\cite{as}.) The fact that the orbit in that case is equivalent to a 
highest weight representation has led to many conflicting arguments
about the entropy of de Sitter spaces. We will have nothing further to
say about this thorny question here.

\section{The Brown-Henneaux ``symmetry'' and the mapping}
A useful way to think about the action (\ref{action2}) is to consider
two separate actions, depending on $g$ and $\bar{g}$ and to impose the
reality condition at the equations of motion. The solutions of
(\ref{action2}) for $g$ can be then seen as holomorphic currents. The
action of $SL(2,C)$ which brings one solution of (\ref{action4}) into
another is:
\be
\begin{array}{c}
g\rightarrow e^{i\log (\partial w)\gamma_0}e^{\frac{i\ell}{2}\partial
\log (\partial w)\gamma_-}g \\
\\
z \rightarrow w(z)
\end{array}
\label{bh}
\ee
One can see directly from (\ref{fixing2}) that this maintains the
K\"ahler structure of the spacial slices at $\scm$. Since it amounts
to a would-be ``pure gauge'' transformation, it can be realized by a
coordinate transformation, which involves redefining the time function
$t$ to arrive at another metric which asymptotes to the de Sitter
metric at $\scm$ \cite{cc}. The current coming from the gauge-fixed
$g$ is: 
\be
J_L=-\partial g g^{-1}dz =
\frac{i}{\ell}\gamma_+dz+\frac{i}{2k\ell}L(z)\gamma_- dz
\label{current}
\ee
So the action (\ref{bh}) is to transform $L(z)$ into
\be
\tilde{L}(w)=\left(\frac{\partial w}{\partial z}\right)^{-2}
\left[L(z)-\frac{k}{4}\{w ; z\}\right]
\ee
We recognize above the anomalous transformation law of the stress
tensor, with central charge $3k=\frac{3\ell}{2 G}$. One then sees the
reason for picking (\ref{fixing3}): it cancels the 
factor of the current proportional to $\gamma_0$ and then maintains
the former on-shell condition (\ref{onshell}).

It is also transparent from the previous discussion that, if one is
able to stick to the gauge choice (\ref{gauge}) throughout time
evolution, the single Liouville mode found above will give rise to a
globally defined metric. For $L(z)$ defined on a sphere with a two
single poles, these solutions have been discussed before \cite{cc,
  bdbm} and have been found to be the Kerr-de Sitter class. The
Brown-Henneaux transformation discussed above generates all such
solutions for meromorphic coordinate transformations. Even more
generically, one can create higher genus space slices in this manner.

However, the effect of the Liouville field in the metric is not felt
until one leaves $\scm$, and in the bulk of space-time any metric is
locally gauge equivalent to the trivial metric. Then, if one is to
assign a gauge-invariant meaning to the Liouville field, one is forced
to see a profile of it as just some parametrization of the conserved
charges of the reduced system, which includes the holonomies of the
gauge field as its simple poles. Echoing the study of Riemann
surfaces, it is the uniformizing coordinate which encodes the
properties of the spatial slices. Being a coordinate choice, it only
fails to be a well-defined gauge transformation because it lives
naturally at the boundary. 

If we consider as an example the Kerr-de Sitter class, one could have
performed the same reduction as in last section for $\scp$. One would
then find another Liouville profile there, which would encode the same
conserved charges. In particular, the positions of the simple poles for
both Liouville modes would be related by a global $SL(2,C)$
transformation. As such the Liouville profiles are essentially
equivalent, amounting just to a global frame in which the global
quantities -- like mass and angular momentum -- are measured. This
is realized in the discussion of last section by the fact that, if one
is able to fix the gauge (\ref{gauge}) throughout the evolution of
space-time, the gauge connection, and consequently the metric, would
be globally written as $A=-d{\cal G}{\cal G}^{-1}$, with ${\cal
  G}=e^{-i\ell^{-1}   t} g(z,\bar{z})$, and then the reduced action
(\ref{action2}) would be the same for both pieces of the boundary.

For generic configurations, the requisition that the gauge is fixed
once and for all seems excessive. Stable causality implies only
that there is a globally well-defined time function, in fact it is
equivalent to it, but it does not imply
that this time function is such that its vector field is geodesic and
everywhere normalized, which are the hidden assumptions behind
(\ref{onshell}) and (\ref{gauge}). Topologically, one can see time
evolution as a cobordism, and this time function as a Morse
function. The mere existence of this Morse function then implies that
the degrees and number of poles in the Liouville field, as well as
their residues, have to be conserved during time evolution. When this
happens, we will say that the Liouville profiles at $\scm$ and $\scp$
are in the same homotopic class. This is how the intuition of
separated worldlines of point particles is imprinted in the Liouville
fields. Generically, lifting this condition would give rise to time
evolutions where the particles could join or split as time progresses,
although one would like at least to enforce conservation 
of total mass and angular momentum to talk about reasonable
spaces. In fact, in this case the condition to enforce on the
space-time seems to be strong causality, for splittings/joinings which
involve small enough point masses would still respect the condition
that there are no closed time-like curves. At any rate, the point here
is that the relation between past and future Liouville profiles is
tied to the time evolution of the system. Solutions which have been
considered so far have the property that time evolution is geodesic and
normalized. They correspond to equal profiles of the Liouville field
at $\scm$ and at $\scp$. 

One can get an insight on what are these solutions in which the
Liouville profiles are not equivalent but are in the same homotopic
class by considering boosted sources. ``Static'' solutions like the
Kerr-de Sitter have the property that the matter current found via the
equations of motion $F={^*J}$ is an $sl(2,C)$-valued one-form which is
zero almost everywhere, but otherwise orthogonal to $e^0_a$. Since $J$
is basically defined to be the Lorentz dual of the stress energy
tensor, this latter fact confirms our intuition of non-interacting
point particles, or, in other words, a configuration of dust. However,
if one makes a $SL(2,C)$ transformation in the $i$-th of those
particles, $J_i \rightarrow U  J_i U^{-1}$, the current may no longer
be orthogonal to $e^0_a$, or even its transformed counterpart. This
comes about because the triad depends on $A$ and it complex conjugate,
so its transformation will depend on $U$ and its complex conjugate,
whereas the matter current transforms by $U$ alone. 

Now consider the quantity $e^0\wedge de^0$. By some manipulation one
can relate it to $e^0 \wedge {^* J^0}$:
\begin{eqnarray*}
e^0\wedge d e^0 & = & \ell \Im ( e^0 \wedge d A^0 ) \\
 & = & - \ell \Im ( e^0 \wedge A^1\wedge A^2 ) + \ell \Im (e^0 \wedge
 {^*J^0}) \\
 & = & - e^0 \wedge e^1 \wedge \omega^2 - e^0 \wedge \omega^2 \wedge
 e^1 + \ell \Im (e^0 \wedge {^*J^0}) \\
 & = & - e^0 \wedge d e^0 + \ell e^0 \wedge \Im({^*J^0})
\end{eqnarray*}
in which we used the definition of the connection in the first and
third lines and the equations of motion in the second. The fourth line
is the condition of zero torsion. So $e^0\wedge de^0 = \frac{\ell}{2}
\Im ( e^0 \wedge {^*J^0})$ and then it is zero when the matter current
corresponds to a distribution of dust. As we discussed above,
generically this will not be the case and this quantity will be non-zero.

The quantity $e^0 \wedge de^0$ measures the failure of the form $e^0$
to be hypersurface orthogonal. By Frobenius' theorem, this means,
among other things, that the space-time does not allow a foliation in
which the vector field $(e^0)^a$ is everywhere orthogonal to the
hypersurfaces. The latter fact forbids us to choose the gauge
(\ref{gauge}) globally. So boosted sources solutions will not be
described by a single Liouville mode, but rather both field profiles, 
at $\scm$ and $\scp$, will be needed to reconstruct the
space-time. Because asymptotically all matter behaves as dust, the
gauge choice is appropriate for generic matter configurations only at
$\scm$ and $\scp$. Asymptotically, the description in terms of the
Liouville fields found in the preceding section will be valid.

One notes that, if the dS/CFT correspondence is implemented in same
manner as in AdS spaces, the space-time information is recovered from
the holographic theory via the renormalization group flow. These are
first order in the scale parameter and hence are determined by the
initial (UV) value of the couplings. By the de Sitter version of the
UV/IR correspondence, this initial value would be naively set at
either $\scp$ or $\scm$. The question that poses itself now is: how
exactly is one supposed to tell the static from the boosted solutions
found in the above discussion if a single Liouville field is blind to
them? The aforementioned argument points to the fact that there are
{\em two} sectors in the holographic theory, which are {\em
  independent} in the UV limit, but whose RG flow induces
correlations. One could think of the two Liouville fields found above
as parametrizing the (space-time) Virasoro current sector of the full
quantum theory, its UV fixed point corresponding to both $\scp$ and
$\scm$. The solutions considered so far, in which the gauge fixing
(\ref{gauge}) is valid globally correspond to ``diagonal'' states
which are the tensor product of equivalent states in each sector of
the holographic theory. Although the particular field representation
of the asymptotic symmetries will change with different dimensions, it
seems plausible to expect that these generic considerations will still
be valid.

\section{The Role of cosmological singularities}

In the last section we showed that there are space-time solutions
corresponding to different states in the sectors $\scm$ and
$\scp$. However, this is not how the known example of holography is
implemented. In fact, in AdS/CFT the UV fixed point is obtained by a
limiting process where degrees of freedom in the interior of
AdS are integrated out. The limiting process effectively gives the
asymptotic value of the fields near, say, $\scm$ and then there is
enough information to reconstruct the whole of space-time. This, of 
course, assumes that time evolution can indeed provide the form of the
field in the bulk of space. 

This assumption was hidden in the preceding discussion in the gauge
choice (\ref{gauge}). As we saw in the preceding section, the
assumption amounts to the constraints between the past and future
profiles discussed in the last session, as dictated by stable
causality. Extending on what has been discussed in last section, let
us consider the cases the spatial slices can undergo topology
changing. We can have disassociation of punctures, as in a single
puncture dividing into two or more, or even the formation of handles
\cite{geons} and generation of non-connected components
\cite{dsfrag}. In the case of disassociation and joining of punctures,
one is naturally led to associate them with Liouville profiles with
different number of poles, 
but with the same total residue to preserve the conserved charges. As
in flat space \cite{gott}, there is a danger of clashing punctures
creating closed time-like curves and thus violating strong
causality. We will see that, also paralleling the flat space scenario
\cite{djth}, there is a mass bound for these types of processes over
which a cosmological singularity will be generated. One can then
obtain the ``big bang'' scenario by a time reversal.

We can see how the classical process of cosmological collapse evolves
by considering the initial value problem. The matter considered will
be an arbitrary distribution of stationary dust $\rho
(z,\bar{z})$. This can be seen as a distribution of 
punctures, arranged so that the total mass is finite. The calculation
is similar to the one done for AdS \cite{Welling}. As we saw in the
second section, the time evolution near $\scm$ is given by a pure
contraction: 
\be
\nabla_a\xi_b=-\frac{1}{\ell}(g_{ab}+\xi_a\xi_b)\mbox{   at }\scm .
\ee
So the natural Ansatz for the solution is:
\be
ds^2=-dt^2+[a(t)]^2 h_{ab}dx^a dx^b.
\ee
Where $h_{ab}$ is a Riemannian two dimensional metric. In the language
of the second section, it is the induced unphysical spatial metric. In
these variables the equations of motion read: 
\be
\begin{array}{rcl}
\displaystyle
 8\pi G a^2
\sqrt{h}\rho - \frac{1}{2}
\sqrt{h}\;^{(2)}\! R & = & -
\bigl(\frac{1}{\ell^2}a^2-\dot{a}^2\bigl) \sqrt{h} \\
\ddot{a}-\frac{1}{\ell^2} a & = & 0 \\
\end{array}
\ee
The second equation means that the left hand side of the first
equation is a constant of motion. Calling $\beta=\ell^{-2}a^2-\dot{a}^2$
and integrating the first equation over the spatial coordinates, we
find: 
\be
\beta = \frac{\pi}{\Sigma}(\chi - 8 G M )\label{bound}
\ee
where $\Sigma$ is the area of the spatial slice in the
unphysical metric and $M$ is
the total mass of the dust configuration. $\chi$ is the
Euler-Poincar\'e characteristic of the spatial slices, which is 2 for
the sphere. One sees that, when $4 G M > 1$, $\beta$ is
negative. When that happens, $a$ is given by:
\be
a(t)=\sinh\left(\frac{t-t_0}{\ell}\right)\label{size}
\ee
with $t_0$ depending on $M$. Hence the ``size of the universe''
vanishes at some finite global time.

This shows that there are states in the boundary theory which
correspond to space-times with cosmological singularities. It is
interesting to note that this bound resembles the bound of masses
which corresponds to operators in the CFT with real scaling weight
\cite{strominger}. For fields whose mass is greater than the de Sitter
mass, the natural Compton wavelength is smaller than the de Sitter
length, so the dust approximation is expected to hold well at large
times then since then interactions and movements will be swamped by
the cosmological constant. At small distances when the pressure begins
to be non-negligible, one can resort to the study of the Raychaudhuri
equation to predict the appearance of the singularity. In fact, a
simple calculation shows that when the energy distribution of a
perfect fluid whose equation of state is $P=w\rho$ satisfies:
\be
8\pi G \rho > \frac{1}{(1-w) \ell^2}
\ee
the expansion parameter will diverge to $-\infty$ at
finite proper time for $w<1$. With all this
in mind, one is then tempted to associate operators in the CFT with 
complex (space-time) scaling weights to space-times with cosmological
singularities \cite{bdbm}. Having no further description of the full
quantum theory, one can only hope that the singularity is resolved
there \cite{bdms}. 

At any rate, the same considerations of last section apply here: the
past (and future) Liouville mode cannot see the difference between
different space-time configurations, as, for instance, those of a gas
of moving particles and the dust configuration above. As in the case
with punctures, details like relative velocities and local
interactions are overwhelmed by the fact that distances between
distinct points are growing exponentially with proper-time. The
question is then how this information is encoded in the boundary
theory. As in the previous section, one is tempted to propose that the
full Hilbert space of the theory is more than just the modes
responsible for the Virasoro currents at $\scm$. If the singularity is
resolved, there may be a way to understand the final state as some
``bounced'' configuration at $\scp$, if the space-time picture is
recovered eventually. 

\section{Whither the Correspondence?}

One of the big puzzles about holography is that it is deeply hidden in
the classical theory, but it is manifest in the working quantum
theories with gravitation we have at the moment. The little we can
infer about its properties is construed from extrapolations of ideas
coming from the study of black holes and (A)dS (semi)-classical
dynamics, like in the preceding discussion. As limited as it is,
general covariance allow us to make some statements about generalities
of any prospective quantum theory.

In the present case, the most curious point is that the particular
quantum theory which gives rise to a de Sitter background ``lives'' in
disjoint regions of space-time. This fact comes about because the
position of the punctures at $\scm$ and $\scp$ are independent. In the
CFT dual language, by the state-operator correspondence, this
means that the state at the $\scm$ may not be the same as that in
$\scp$. In fact, as argued in the last section, one would generically
like to specify those two independent 
states in the dual theory and then ask questions about space-time
behavior. This also has some echoes of the study
of quantum fields in an eternal black hole background \cite{bd} and
more recently in global de Sitter spaces \cite{bms}, where there
is a distinction between the Schwarzschild Fock space and the Kruskal
Fock space. Locality tells us that the former splits into two subsets,
corresponding to the visible and not visible exterior regions of the
extended space-time. The vacuum of the quantum field is a pure state
in the Kruskal (or Hartle-Hawking) Fock space, but expectation values
of local operators in either exterior can only see the projection of
the state into the corresponding Schwarzschild Fock space. Some
arguments have been put forward to extend this point of view to
gravity itself \cite{cc1, malda}. The results presented here point to
the fact that this may be even more generic, in that regions separated
by a horizon, be it of a black hole or of the observable universe, are
described by a theory whose Fock space naturally decomposes into an
enumerable set of subspaces (assuming that such description is indeed
valid in the UV fixed point). In such theory the existence of a
horizon is encoded by correlations between the distinct subspaces,
{\it e.~g.} the ${\cal I}^\pm$ correlators found in \cite{bms},
and locality outside the horizon is translated to the fact that
observable quantities, the ``true observables'' of \cite{bms}, are
obtained by projection onto one particular subspace of the
theory. Quantities which are defined in the whole of space-time are
the meta-observables of \cite{w-ds}. 

In the AdS/CFT case these subtleties could be overlooked in simple
deformations of the background since its holographic screen is
connected, for in global AdS there is no horizon. For dS/CFT, even
in the simplest case, global de Sitter, our na\"{\i}vet\'e is
exposed. In the AdS case, even in the presence of a horizon one could
consistently truncate the full spectrum of the theory into those
states seen by the visible spatial infinity. The horizon is then seen
as an impenetrable membrane \cite{susskind1}. For de Sitter, this line
of reasoning leads to a paradox \cite{susskind2}. The discussion in
this paper infers that the kernel of the problem is in the truncation,
which cannot be done consistently for generic perturbations of the
global de Sitter background. Recently, this fact has been stressed by
R. Bousso {\it et~al.} \cite{bdwm} which tried to carry out such
truncation for theories with interacting form fields. Although such
fields are non-local excitations in the boundary CFT, and in this
sense complementary to what has been done here, both results agree in
casting doubt on the program of setting up a quantum theory
of gravity in de Sitter backgrounds with a finite number of degrees of
freedom \cite{banks}.

As it is customary, there are more questions raised than answers. The
particularly hard question is to understand those CFT states
corresponding to $\beta < 0$ in (\ref{bound}). Or, in other words, how
to make sense of operators with complex scaling dimension in such
CFT. In dS$_3$ one can create such type of background 
by a suitable identification of global de Sitter by an
element of the isometry group. In this case the identification is
obviously singular but this particular type of singularity merited
some attention recently \cite{ckr}. Another, perhaps more amenable,
question deals with the entropy of these backgrounds. As in the AdS
case, the Liouville mode is the simple result of global covariance and
therefore cannot distinguish between quantum states giving rise to the
same geometry. This fact is encoded by semi-classical calculations
showing that $c_{\mbox{eff}}=1$ for the Liouville field \cite{ks}. On
the other hand it is extremely confusing how exactly the would-be
gauge symmetries (the Brown-Henneaux type discussed in section two)
should act on the quantum states. It is unclear whether one can
continue tackling these questions by thinking in terms of classical
gravity alone.

When this manuscript was in its final phase of revision, the work of
Balasubramanian {\it et al.} \cite{bdbm2} appeared in which there is
some overlap with the points discussed in section III and V.

\acknowledgments

I would like to thank specially Emil Martinec and Glic\'eria Carneiro
da Cunha for discussions,
guidance and support during this project. I would also like to
acknowledge Juan Maldacena, Li-Sheng
Tseng, Will McElgin and Andrei Parnachev, whose help was precious
during the time this work was maturing. Geworfenheit.

\hbadness 10000

\end{document}